\documentclass[12pt,a4paper]{article}
\usepackage[leqno]{amsmath}
\title{On the modified method of simplest equation and the nonlinear Schr{\"o}dinger equation} 
\author{ Nikolay K. Vitanov$^1$, Zlatinka I. Dimitrova$^2$} 
\date{ $^1$Institute of Mechanics, Bulgarian Academy of Sciences, Akad. G. Bonchev Str., Bl. 4, 1113 Sofia, Bulgaria\\
$^2$ "G. Nadjakov" Institute of Solid State Physics, Bulgarian Academy of Sciences, Blvd. Tzarigradsko Chaussee 72, 1784 Sofia, Bulgaria}
\begin{document}
\maketitle
\begin{abstract}
We consider an extension of the methodology of the
modified method of simplest equation to the case of
use of two simplest equations. The extended methodology is applied for obtaining exact solutions of model nonlinear partial differential equations for deep water
waves: the nonlinear Schr{\"o}dinger equation. It is shown  that the
methodology works also for other equations of the nonlinear Schr{\"o}dinger kind.
\end{abstract}
\section{Introduction}
In the last decades we have observed a fast growth of research on nonlinear phenomena \cite{np1} - \cite{np5}. Today this is well established research area
and the research on nonlinear waves finds its significant place within this
area. One of the most studied models in the research on waves are 
the models of  water waves.
The nonlinear Schr{\"o}dinger equation (NSE) is a model equation for nonlinear waves in a deep water. Thus the traveling wave solutions of this equation are of large interest for fluid mechanics. Below we discuss the nonlinear
Schr{\"o}dinger equation from the point of view of  the modified method of simplest equation \cite{v1} - \cite{v3} for obtaining exact analytical solutions of nonlinear partial differential equations. The goal is to refine the
existing methodology in order to make this methodology capable to obtain exact   traveling wave solutions of the NSE. 
\par
 Numerous models of natural and social systems are based on differential equations \cite{murr}-\cite{debn}. Often the model equations are nonlinear partial differential equations that
have traveling-wave solutions. These traveling wave 
solutions are studied very intensively  \cite{scott}-\cite{ablowitz1}.
Powerful methods exist for obtaining exact traveling-wave solutions
of nonlinear partial differential equations 
, e.g., the method of inverse scattering transform or the method of Hirota 
\cite{ablowitz2} - \cite{hirota}. These methods work very well for the case
of integrable nonlinear PDEs. Many other approaches for obtaining exact special
solutions of nonintegrable nonlinear PDEs have been developed in the recent years (for examples see \cite{he} - \cite{wazw2}). Below we shall consider
the method of simplest equation  and our focus will be on a version of this
method called modified method of simplest equation \cite{kudr05x}, \cite{kudr08}, \cite{v1} - \cite{v3}.
The method of simplest equation is based on  a procedure analogous to the first 
step of the test for the Painleve property \cite{kudr05x} - \cite{k10}. In the version of the method called  modified method of simplest equation \cite{v1} - \cite{v3} this procedure is substituted 
by  the concept for the balance equation. We are going to discuss below
the possibility of use of more than one simplest equation for obtaining
exact analytical solution of the solved nonlinear partial differential
equation. The modified method of simplest equation has  many applications,
e.g., obtaining exact traveling wave solutions  of generalized Kuramoto - Sivashinsky equation, reaction - diffusion  equation, reaction - telegraph equation\cite{v1}, \cite{vd10},
generalized Swift - Hohenberg equation and generalized Rayleigh equation \cite{v2}, 
generalized Fisher equation, generalized Huxley equation \cite{vit09x}, 
generalized Degasperis - Procesi equation and b-equation \cite{vit11a}, 
extended Korteweg-de Vries equation \cite{v13}, etc. \cite{v13a}, \cite{v14}.
\section{Modified method of simplest equation}
\subsection{The classic version of the methodology}
The methodology of the modified method of simplest equation used up to now is appropriate
for solving model nonlinear PDEs, eg., from the research area of shallow water waves.
These equations contain real functions of a coordinate and of the time and because of
this one simplest equation is sufficient for application of the methodology. The
model equations for the deep water waves contain complex functions. Because of
this we have to extend the methodology of the modified method of simplest equation to
the case of use of two simplest equations. The classic version of the modified method of simplest
equation follows below. The extension of the methodology is presented in the
following subsection.
\par
The  methodology based on the use of one simplest equation is
described, e.g., in \cite{vit11b}. We note that cases of use of more than 
one simpler differential equation for obtaining exact analytical solutions
of more complicated partial differential equation can be found in the
literature even 25 years ago, e.g., \cite{v92a} - \cite{vit96b}.
First of all by means of an 
appropriate ansatz the solved  nonlinear partial differential 
equation is reduced to 
a nonlinear  differential equation, containing derivatives of a function
\begin{equation}\label{ix1}
P \left( u(\xi), u_{\xi},u_{\xi \xi},\dots,\right) = 0
\end{equation}
Then the function $u(\xi)$   is searched as some 
function of another function, e.g., $g(\xi)$, i.e.
\begin{equation}\label{ix1x}
u(\xi) = G[g(\xi)]
\end{equation} 
Note that the kind of the function $F$ is not
prescribed. Often one uses a finite-series relationship, e.g., 
\begin{equation}\label{i2}
u(\xi) = \sum_{\mu_1=-\nu_1}^{\nu_2} p_{\mu_1} [g (\xi)]^{\mu_1}; 
\end{equation}
where $p_{\mu_1}$ are coefficients. The function  $g(\xi)$ 
is solutions of a simpler ordinary differential equation called simplest 
equation. Eqs.(\ref{ix1x}) are substituted in Eq.(\ref{ix1}) and let the result
of this substitution be a polynomial containing $g(\xi)$. 
What follows is application of a balance procedure.  This procedure has to ensure 
that all of the coefficients of the obtained polynomial of $g(\xi)$ 
contain more than one term. The procedure leads to a
balance equation for some of the parameters of the solved equation and
for some of the parameters of the solution. Eqs. (\ref{ix1x})  describes a
candidate for solution of Eq.(\ref{ix1}) if all coefficients of the obtained 
polynomial of are equal to $0$. This condition leads to a system of 
nonlinear algebraic equations for the coefficients of the solved nonlinear PDE and for the
coefficients of the solution. Any nontrivial solution of this algebraic system
leads to a solution the studied  nonlinear partial differential equation.
\subsection{Extension of the methodology to the case of use of two simplest equations}
Now we consider an extension of the methodology to the
case where the idea of use of two simplest equations is incorporated in the methodology.  First of all by means of an 
appropriate ans{\"a}tze (for an example two traveling-wave
ans{\"a}tze) the solved  nonlinear partial differential equation is reduced to 
a nonlinear  differential equation, containing derivatives of two functions
\begin{equation}\label{i1}
P \left( u(\xi), u_{\xi},u_{\xi \xi},\dots, w(\zeta), w_\zeta,
w_{\zeta \zeta}, \dots \right) = 0
\end{equation}
Then the functions $u(\xi)$, $w(\zeta)$,  are searched as some 
function of another functions, e.g., $g(\xi)$ and $\phi(\zeta)$, i.e.
\begin{equation}\label{i1x}
u(\xi) = G[g(\xi)]; \ \ w(\zeta) = F[\phi(\zeta)]; \dots
\end{equation} 
Note that the kind of the functions $F$ and $G$  is not
prescribed. Often one uses a finite-series relationship, e.g., 
\begin{equation}\label{i2}
u(\xi) = \sum_{\mu_1=-\nu_1}^{\nu_2} p_{\mu_1} [g (\xi)]^{\mu_1}; \ \ \ 
w(\zeta) = \sum_{\mu_2=-\nu_3}^{\nu_4} p_{\mu_2} [\phi (\zeta)]^{\mu_2}, 
\end{equation}
where $p_{\mu_1}$, $p_{\mu_2}$ are coefficients. The functions  $g(\xi)$, $\phi(\zeta)$ 
are solutions of simpler ordinary differential equations called simplest 
equations. Eq.(\ref{i1x}) is substituted in Eq.(\ref{i1}) and let the result
of this substitution be a polynomial containing $g(\xi)$ and $\phi(\zeta)$. Then a balance procedure is applied.  This procedure has to ensure 
that all of the coefficients of the obtained polynomial of $g(\xi)$ and $\phi(\zeta)$ contain more than one term. The procedure leads to one or more
balance equations for some of the parameters of the solved equation and
for some of the parameters of the solution. Eqs. (\ref{i1x}) could describe a
candidate for solution of Eq.(\ref{i1}) if all coefficients of the obtained 
polynomial of are equal to $0$. This condition leads to a system of nonlinear algebraic equations for the coefficients of the solved nonlinear PDE and for the
coefficients of the solution. Any nontrivial solution of this algebraic system
leads to a solution the studied  nonlinear partial differential equation.
\par 
The organization of the text below is as follows. In Sect. 2 we discuss an
application of the  extension of the modified method of simplest equation described above  to the classic nonlinear Schr{\"o}dinger equation and to another
equation of the same class. Several concluding remarks are summarized in
Sect. 3.
\section{Application to nonlinear Schr{\"o}dinger equation and similar equations}
The classic nonlinear Schr{\"o}dinger equation is
\begin{equation}\label{cls}
i q_t +  a q_{xx} + b q \mid q \mid^2 =0
\end{equation}
where $i=\sqrt{-1}$ and $a$ and $b$ are parameters. The solution of Eq.(\ref{cls}) will be searched as
\begin{equation}\label{eq2}
q(x,t) = g(\xi) h(x,t)
\end{equation}
where $g(\xi)$ is a real function ($\xi = \alpha x + \beta t$) and $h(x,t)$ is a complex function.
The two simplest equations will be for the functions $g(\xi)$ and $h(x,t)$ respectively. Let us denote
as $h^*(x,t)$ the complex conjugate function of $h(x,t)$.  The substitution of Eq.(\ref{eq2}) in
Eq.(\ref{cls}) leads to (we denote $g_\xi$ as $g'$)
\begin{equation}\label{eq3}
i \beta g'h + igh_t + \alpha^2 a g'' h + 2 \alpha a g'h_x + agh_{xx} + 
b g^3 h^2 h^* = 0 
\end{equation}
\par
The first simplest equation is for the function $h(x,t)$. Taking into an account
the presence of $h$ and its derivatives in Eq.(\ref{eq3}) as well as the
presence of the term $hh^*$ there and aiming to choose such simplest equation that will lead to reduction of Eq.(\ref{eq3}) to an equation for 
$g(\xi)$ we arrive at the simplest equation
\begin{equation}\label{eq4}
h_\zeta = i h, \ \ \zeta = \kappa x + \omega t + \theta 
\end{equation}
which solution is
\begin{equation}\label{eq5}
h(\zeta) = \exp(i \zeta) =  \exp[i(\kappa x + \omega t + \theta)]
\end{equation}
The substitution of Eq.(\ref{eq4}) in Eq.(\ref{eq3}) reduces  Eq.(\ref{eq3}) to
an equation for the function $g(\xi)$
\begin{equation}\label{eq6}
\alpha^2 a g'' + (2\alpha \kappa a + \beta)ig' - (\omega + \kappa^2 a)g + b g^3  =0
\end{equation}
$g(\xi)$ has to be a real function and then
\begin{equation}\label{eq7}
\beta = - 2 \alpha \kappa a
\end{equation}
The substitution of Eq.(\ref{eq7}) in Eq.(\ref{eq6}) followed by multiplication
of the result by $g'$ and integration with respect to $\xi$ leads to the
equation
\begin{equation}\label{eq8}
\alpha^2 a g'^2 - (\omega + \kappa^2 a)g^2 - c + \frac{bg^4}{2}=0
\end{equation}
where $c$ is a constant of integration. Further we set
\begin{equation}\label{eq9}
u = g^\sigma
\end{equation}
where $\sigma$ is a parameter that will be determined below. The substitution of
Eq.(\ref{eq9}) in Eq.(\ref{eq8}) leads to the following equation for $u(\xi)$
\begin{equation}\label{eq10}
u'^2 = \frac{\sigma^2(\omega+\kappa^2a)}{\alpha^2 a} u^2 + \frac{\sigma^2 c}{
\alpha^2 a} u^{2(\sigma-1)/\sigma} -  \frac{b \sigma^2}{2 \alpha^2 a } u^{2(1+\sigma)/\sigma}
\end{equation}
Two cases are possible here $\sigma=1$ and $\sigma=2$.
For the case $\sigma=1$ Eq.(\ref{eq10}) becomes
\begin{equation}\label{eq11}
u'^2 =\frac{ c}{\alpha^2 a} + \frac{(\omega+\kappa^2a)}{\alpha^2 a} u^2  -  \frac{b}{2 \alpha^2 a } u^4
\end{equation}
This
equation contains as particular case the equation for the elliptic functions of Jacobi
\begin{equation}\label{jacobi}
u'^2=p + q u^2 + r u^4
\end{equation}
(for ${\rm sn}(\xi;k^*)$: $p=1,q=-(1+{k^*}^2),r={k^*}^2$; for ${\rm cn}(\xi;k^*)$: $p=1-{k^*}^2,q=2{k^*}^2-1,r=-{k^*}^2$;
for ${\rm dn}(\xi;k^*)$: $p=-(1-{k^*}^2),q=2-{k^*}^2,r=-1$; $k^*$ is the modulus of the corresponding elliptic function of Jacobi).
\par
For the case $\sigma=2$ Eq.(\ref{eq10}) becomes
\begin{equation}\label{eq12}
u'^2 = \frac{4c}{\alpha^2 a}u +\frac{4(\omega+\kappa^2a)}{\alpha^2a}u^2 - \frac{2b}{\alpha^2a}u^3
\end{equation}
This equation contains as particular case the equation for the elliptic functions of Weierstrass $\wp(\xi, \gamma_2, \gamma_3)$:
\begin{equation}\label{wp}
\wp'^2 = 4 \wp^3 - \gamma_2 \wp - \gamma_3
\end{equation} 
where $\gamma_2$ and $\gamma_3$ are parameters.
\par
Let us first consider Eq.(\ref{eq11}) and write a solution on the basis of
the Jacobi elliptic function $\rm{cn}(\xi,k^*)$. For this case we have to solve
the system of equations
\begin{eqnarray}\label{eq13}
\hskip1cm
p=\frac{c}{\alpha^2 a} = 1 - {k^*}^2; \ \ 
q = \frac{\omega + \kappa^2 a}{\alpha^2 a} = 2 {k^*}^2 - 1 ; \ \
r = - \frac{b}{2 \alpha^2 a} = - {k^*}^2
\end{eqnarray}
The solution of this system is
\begin{equation}\label{eq14}
\alpha^2 = \frac{b}{2a{k^*}^2}; \ \ 
c = \frac{b(1-{k^*}^2)}{2 {k^*}^2}; \ \
\omega = \alpha^2 a (2{k^*}^2 - 1) - \kappa^2 a
\end{equation}
and the corresponding solution of the nonlinear Schr{\"o}dinger equation 
(\ref{cls}) is
\begin{eqnarray}\label{eq15}
q(x,t)= {\rm cn} \Bigg[ \left( \frac{b}{2a {k^*}^2}\right)^{1/2} x + \beta t; k^* \Bigg] \times \nonumber \\\exp \Bigg[ i \Bigg(\kappa x +  a[\alpha^2  (2{k^*}^2 - 1) - \kappa^2]t + \theta \Bigg) \Bigg]
\end{eqnarray}
When $k^*=1$ then the Jacobi elliptic function ${\rm cn}$ is reduced to ${\rm sech} (\xi)$. In this case the solution (\ref{eq15}) becomes
\begin{eqnarray}\label{eq16}
q(x,t)= {\rm sech} \Bigg[ \left( \frac{b}{2a }\right)^{1/2} x + \beta t \Bigg] \times \nonumber \\\exp \Bigg[ i \Bigg(\kappa x +  a[\alpha^2  - \kappa^2]t + \theta \Bigg) \Bigg]
\end{eqnarray}
\par
Let us now consider (\ref{eq12}) for the particular case where the
solution can be obtained on the  basis of the elliptic function of Weierstrass
(\ref{wp}). In this case $\gamma_3 = 0$; $\alpha = \pm[-b/(2a)]^{1/2}$; $\omega = -\kappa^2 a$. Thus the corresponding solution of the nonlinear equation of
Schr{\"o}dinger becomes
\begin{eqnarray}\label{eq17}
q(x,t) = \wp \Bigg \{ \Bigg [ \pm \Bigg( -\frac{b}{2a} \Bigg)^{1/2}  (x- 2 \kappa a)\Bigg ] ; \frac{8c}{b}, 0 \Bigg \}^{1/2} \exp [i(\kappa x  - \kappa^2 a t + \theta)] \nonumber \\
\end{eqnarray}
\subsection{Another example for application of the extended methodology}
Let us now consider more complicated equation of the nonlinear Schr{\"o}dinger kind, i.e.,
\begin{equation}\label{eq18}
iq_t + a q_{xx} + b_{-2} q \mid q \mid^{-4} + b_0 q + b_1 q \mid q \mid^2 +
b_2 q \mid q \mid^4 =0
\end{equation}
where $b_i$ are parameters. The application of the extended version of the
modified method of simplest equation leads to Eqs. (\ref{eq4}), (\ref{eq5})
for $h(\zeta)$ and for $g(\xi)$ we obtain $g=u^{1/2}$ associated with
the following (simplest) equation
\begin{equation}\label{eq19}
u'^2 = c_{-2} + c_0 u^2 + c_1 u^3 + c_2 u^4
\end{equation}
where
$$
c_{-2} = \frac{4b_{-2}}{\alpha^2 a}; \ c_0 = \frac{4(\omega + \kappa^2 a - b_0)}{\alpha^2 a}; \ c_k = - \frac{4 b_k}{\alpha^2 a(k+1)}, \ k=1,2
$$
We shall consider the case $c_1=0$ and then Eq.(\ref{eq19}) is reduced to the
differential equation for the elliptic functions of Jacobi. Let us write
the solution based on the Jacobi elliptic function ${\rm sn}(\xi, k^*)$. $k^*$
is the modulus of the Jacobi elliptic function and for the case of the ${\rm
sn}$ function we have the following relationships among the modulus of the
elliptic function and the coefficients in Eq.(\ref{eq19})
$$
c_{-2} =1; \ c_{0} = -(1+{k^*}^2); \ c_2 = {k^*}^2
$$ 
Then
\begin{equation}\label{eq20}
{k^*}^2 = \frac{b_2}{3 b_{-2}}; \ \alpha^2 = \frac{4 b_{-2}}{a}; \
\omega = b_0 - \kappa^2 a - \frac{1}{3}(b_2 + 3 b_{-2})
\end{equation}
Thus the searched solution is
\begin{eqnarray}\label{eq21}
q(x,t) = \left \{{\rm sn}\left[ \left( \frac{4b_{-2}}{a}\right)^{1/2}(x - 2 \kappa a t);k^* \right] \right\}^{1/2}\exp \Bigg \{ i \Bigg[ \kappa x + \Bigg(b_0 - \kappa^2 a - \nonumber \\
 \frac{1}{3}(b_2 + 3 b_{-2}) \Bigg) t + \theta)\Bigg ] \Bigg \}
\end{eqnarray}
\section{Discussion}
As we have seen above the methodology of the modified method of simplest
equation is effective tool for obtaining exact particular solutions of
nonlinear partial differential equations. The emphasis of this paper was on
the extension of the methodology in order to include the model equations for
deep water waves in the class of equation that can be treated by the
modified method of simplest equation. In order to do this we have to
use more than one simplest equation. We note that the extended methodology 
can be applied to many equations of nonlinear Schr{\"o}dinger kind. It is known that numerous equation of NSE kind have solitary wave solutions \cite{b1} - \cite{p7}. Thus the perspectives are very good for obtaining new solutions of these model equation and for application of the
extended methodology to more complicated equations from the area of water waves and optics. 
\par
Of course the methodology can be extended to the case of use of more than two simplest equations simultaneously. This research and its results will be reported elsewhere.


\begin{thebibliography}{99}

\bibitem{np1}
HAKEN, H. Advanced Sinergetics. Instability Hierarchies of Self-Organizing Systems and Devices. Springer, Berlin, 1983.
\bibitem{puu}
PUU, T. Attractors, Bifurcations, \& Chaos: Nonlinear Phenomena in Economics. Springer, Bearlin, 2003.
\bibitem{wigen}
WIGEN, P. E. (Ed.) Nonlinear Phenomena And Chaos In Magnetic Materials. World Scientific, Singapore, 1994.
\bibitem{grz}
GRZYBOWSKI, B. A. Chemistry in motion: Reaction-diffusion systems for micro- and nanotechnology. Wiley, Chichester, 2009. 
\bibitem{matra}
MARASULOV, D., H. E. STANLEY. Nonlinear Phenomena in Complex Systems: From Nano to Macro Scale. Springer, Dordrecht, 2013.
\bibitem{np2}
BAKUNIN, O. G. Turbulence and Diffusion. Scaling versus Equations. Springer, Berlin, 2008.
\bibitem{takeuti}
TAKEUTI, M., J. R. BUCHLER. (Eds.) Nonlinear Phenomena in Stellar Variability. Springer, Dordrecht, 1993.
\bibitem{cordero}
CORDERO, P., B. NACHTERGAELE. Nonlinear Phenomena in Fluids, Solids and other Complex Systems. North-Holland, Amsterdam, 1990.
\bibitem{boeck}
BOECK, T., N. K. VITANOV. Low-dimensional chaos in zero-Prandtl-number Benard–Marangoni convection. \emph{Phys. Rev. E} \textbf{65} (2002), 037203.
\bibitem{chab}
CHABCHOUB, A., N. VITANOV, N. HOFFMANN. Experimental evidence for breather type dynamics in freak waves. \emph{PAMM}, \textbf{10}, 2010, 495 - 496.
\bibitem{np3}
HAKEN, H. Brain Dynamics. An Introduction to Models and Simulations. Springer, Berlin, 2008.
\bibitem{banerjee}
BANERJEE, S., G. C. Verghese. Nonlinear Phenomena in Power Electronics: Bifurcations, Chaos, Control, and Applications.
Wiley, New York, 2001.
\bibitem{k1}
KANTZ, H., D. HOLSTEIN, M. RAGWITZ, N. K. VITANOV. Markov chain model for turbulent wind speed data. \emph{Physica A}
\textbf{342} (2004), 315 - 321.
\bibitem{panch}
PANCHEV, S., T. SPASSOVA, N. K. VITANOV. Analytical and numerical investigation of two families of Lorenz-like dynamical systems. \emph{Chaos, Solitons \& Fractals},
\textbf{33} (2007), 1658 - 1671.
\bibitem{nikolova}
PETROV, V., E. NIKOLOVA, O. WOLKENHAUER. Reduction of nonlinear dynamic systems with an application to signal transduction pathways. \emph{IET Systems Biology}, \textbf{1} (2007), 2 - 9.
\bibitem{nikolova2}
NIKOLOVA, E. New result in Ras/Raf/MEK/ERK signal pathway dynamical model. \emph{COMPTES RENDUS-ACADEMIE BULGARE DES SCIENCES} \textbf{59}, 143 - 150.
\bibitem{mosekilde}
MOSEKILDE, E., O. G. MOURITSEN. Modelling the Dynamics of Biological Systems: Nonlinear Phenomena and Pattern Formation.
Springer, Berlin, 1995.
\bibitem{d1}
DIMITROVA, Z. I., N. K. VITANOV. Influence of adaptation on the nonlinear dynamics of a system of competing populations.
\emph{Physics Letters A} \textbf{272} (2000) 368 - 380.
\bibitem{leonov}
LEONOV, A. I., A. N. PROKUNIN. Nonlinear Phenomena in Flows of Viscoelastic Polymer Fluids. Springer, Dordrecht, 1994.
\bibitem{cantrell}
CANTRELL, R.S., C. COSTNER. Spatial ecology via reaction-diffusion equations. Wiley, Chichester, 2003. 
\bibitem{d3}
DIMITROVA, Z. I., N. K. VITANOV. Chaotic pairwise competition. \emph{Theoretical Population Biology} \textbf{66} (2004), 1 - 12.
\bibitem{np4}
SORNETTE, D. Critical Phenomena in Natural Sciences. Springer, Berlin, 2006.
\bibitem{d2}
DIMITROVA, Z. I., N. K. VITANOV. Adaptation and its impact on the dynamics of a system of three competing populations. |emph{Physica A} \textbf{300} (2001), 91 - 115.
\bibitem{alg}
AL-GHOUL, M., B. C. EU. Hyperbolic reaction-diffusion equations, patterns, and phase speeds for the Brusselator.
\emph{Journal of Physical Chemistry}, \textbf{100} (1996) 18900 -- 18910.
\bibitem{np5}
FRANK, T. D. Nonlinear Fokker-Planck Equations. Fundamentals and Applications.
Springer, Berlin, 2005.
\bibitem{v1} 
	VITANOV, N.K., Z.I DIMITROVA , H. KANTZ. Modified Method of Simplest Equation and its Application to Nonlinear PDEs. \emph{Applied Mathematics and Computation} \textbf{216} (2010),  
2587 -- 2595.
\bibitem{v2} 
	VITANOV, N.K. Modified Method of Simplest Equation: Powerful Tool for Obtaining Exact and Approximate Traveling-Wave Solutions of Nonlinear PDEs. \emph{Communications in Nonlinear Science and  Numerical Simulation} \textbf{16} (2011), 16: 1176 -- 1185.
\bibitem{v3} 
	VITANOV N.K.,  Z. I. DIMITROVA, K. N. VITANOV. Modified Method of Simplest Equation for Obtaining Exact Analytical Solutions of Nonlinear Partial Differential Equations: Further Development of the Methodology with Applications. \emph{Applied Mathematics and Computation} \textbf{269} (2015),  363 -- 378.
\bibitem{murr}
	MURRAY J.D. Lectures on Nonlinear Differential Equation Models in Biology. Oxford University Press, Oxford,  1977.
\bibitem{ames}
AMES, W. F. Nonlinear partial differential equations in engineering. Academic Press, New York, 1972.
\bibitem{ac}
	ABLOWITZ M. , P. A. CLARKSON. Solitons, Nonlinear Evolution Equations and Inverse Scattering. Cambridge University Press, Cambridge, 1991.
\bibitem{benkirane}
BENKIRANE, A., J.-P. GOSSEZ (Eds.) Nonlinear patial differential equations. Addison Wesley Longman, Essex, UK, 1996.
\bibitem{knvit}
	VITANOV N. K. Science Dynamics and Research Production. Indicators, Indexes, Statistical Laws and Mathematical Models.  Springer, Cham,  2016.
\bibitem{vsd2}
	VITANOV N. K., K. N. VITANOV. Box Model of Migration Channels. \emph{Mathematical
	Social Sciences} \textbf{80} (2016), 108 -- 114.
\bibitem{debn}
	DEBNATH L.  Nonlinear  Partial Differential Equations for Scientists and Engineers. Springer, New York, 2012.
\bibitem{scott}
	SCOTT A. C.  Nonlinear  Science. Emergence and Dynamics of Coherent Structures. Oxford University Press, Oxford, 1999.
\bibitem{fan}
FAN, E., Y. C. HON. A series of travelling wave solutions for two variant Boussinesq equations in shallow water waves. \emph{Chaos, Solitons \& Fractals} \textbf{15} (2003) 559 -- 566.
\bibitem{vit09a}
	VITANOV N.K., I. P. JORDANOV, Z. I. DIMITROVA. On Nonlinear Population Waves.\emph{
	Applied Mathematics and Computation}. \textbf{215} (2009), 2950 -- 2964.	
\bibitem{holmesx}
	HOLMES P., J. L. LUMLEY, G. BERKOOZ. Turbulence, Coherent Structures, Dynamical Systems and Symmetry. Cambridge University Press, Cambridge, 1996.	
\bibitem{gallay}
GALLAY, T.,R. RAUGEL. Scaling variables and stability of hyperbolic fronts. \emph{SIAM J. Math. Anal.} \textbf{32} (2000) 1 -- 29.
\bibitem{tabor}
        TABOR M. Chaos and Integrability in Dynamical Systems. Wiley, New York,  1989.	
\bibitem{vit09}
	VITANOV N.K., I. P. JORDANOV, Z. I. DIMITROVA. On Nonlinear Dynamics of Interacting Populations: Coupled Kink Waves in a System of Two
	Populations. \emph{Commun. Nonlinear Sci. Numer. Simulat}.  \textbf{14} (2009), 2379 -- 2388.
\bibitem{ablowitz1}
	ABLOWITZ M. J., D. J. KAUP, A. C. NEWELL. Nonlinear Evolution Equations of Physical Significance. \emph{Phys. Rev. Lett.} \textbf{31} (1973), 125 -- 127.	
\bibitem{ablowitz2}
	ABLOWITZ M. J.,  D. J. KAUP, A. C. NEWELL, H. SEGUR. Inverse Scattering
	Transform - Fourier Analysis for Nonlinear Problems. \emph{Studies in
	Applied Mathematics} \textbf{53} (1974),  249 -- 315.
\bibitem{gardner}
GARDNER, C. S., J. M. GREENE, M. D. KRUSKAL,  R. R. MIURA. Method for solving Korteweg- de Vries equation. \emph{Phys. Rev. Lett.} \textbf{19} (1967) 1095 -- 1097.
\bibitem{hirota}
	HIROTA R. Exact Solution of Korteweg-de Vries Equation for
	Multiple Collisions of Solitons. \emph{Phys. Rev. Lett.} \textbf{27} (1971),  1192 -- 1194.
\bibitem{he}
	HE, J.-H., X. -H. WU. Exp-Function Method for Nonlinear Wave Equations.
	\emph{Chaos, Solitons \& Fractals} \textbf{30} (2006)  700 -- 708.	
\bibitem{herrem}
	MALFLIET W.,  W. HEREMAN. The Tanh Method: I. Exact Solutions of Nonlinear 	
	Evolution and Wave equations. \emph{Physica Scripta} \textbf{54} (1996);  563 -- 568.
\bibitem{wazw2}
	WAZWAZ, A.-M. Partial Differential Equations and Solitary Waves Theory.
	Springer, Dordrecht,  2009.
\bibitem{kudr05x}
	KUDRYASHOV N. A. Simplest Equation Method to look for Exact Solutions
	of Nonlinear Differential Equations. \emph{Chaos Solitons \& Fractals} \textbf{24} ()2005),
	1217 -- 1231.
\bibitem{kudr08}
	KUDRYASHOV N. A., N. B. LOGUINOVA. Extended Simplest Equation Method
	for Nonlinear Differential Equations. \emph{Applied Mathematics and Computation} \textbf{205} (2008), 396 -- 402.
\bibitem{kudr05}
	KUDRYASHOV N. A. Exact Solitary waves of the Fisher Equation. \emph{Phys. Lett. A} \textbf{342} (2005),  99 -- 106.	
\bibitem{k10}
	KUDRYASHOV N. A. Meromorphic Solutions of Nonlinear Ordinary Differential Equations. \emph{Communications in Nonlinear Science and Numerical Simulation} \textbf{15} (2010),  2778 -- 2790.
\bibitem{vd10}
	VITANOV N. K., Z. I. DIMITROVA. Application of the Method of Simplest Equation for Obtaining Exact Traveling-Wave Solutions for Two Classes of Model PDEs from Ecology and Population Dynamics. \emph{ Commun. Nonlinear Sci.  Numer. Simulat.} \textbf{15} (2010), 2836 -- 2845.	
\bibitem{vit09x}
	VITANOV N. K. Application of Simplest Equations of Bernoulli and Riccati Kind for Obtaining Exact Traveling Wave Solutions for a Class of PDEs with
	Polynomial Nonlinearity. \emph{Commun. Nonlinear Sci.  Numer. Simulat.} \textbf{15} (2010), 2050 -- 2060.
\bibitem{vit11a}
	VITANOV N. K., Z. I. DIMITROVA, K. N. VITANOV. On the Class of Nonlinear PDEs that can be treated by the Modified Method of Simplest Equation. Application to Generalized Degasperis - Processi Equation and B-equation. \emph{Commun. Nonlinear Sci. Numer. Simulat.} \textbf{16} (2011) 3033 -- 3044.
\bibitem{v13}
	VITANOV N. K., Z. I. DIMITROVA, H. KANTZ. Application of the Method of Simplest Equation for Obtaining Exact Traveling-Wave Solutions for the Extended Korteweg-de Vries Equation and Generalized Camassa-Holm Equation. \emph{Applied Mathematics and Computation} \textbf{219} (2013), 7480 -- 7492.
\bibitem{v13a}
	VITANOV N. K., Z. I. DIMITROVA, K. N. VITANOV. Traveling Waves and Statistical 
	Distributions Connected to Systems of Interacting Populations. \emph{Computers \& Mathematics with Applications} \textbf{66} (2013),  1666 -- 1684.
\bibitem{v14} 
	VITANOV N. K., Z. I. DIMITROVA. Solitary Wave Solutions for
	Nonlinear Partial Differential Equations that contain Monomials of Odd and 
	Even Grades with respect to Participating derivatives. \emph{Applied Mathematics and
	Computation} \textbf{247} (2014), 213 -- 217.
\bibitem{vit11b}
    VITANOV N. K. On Modified Method of Simplest Equation for Obtaining 
	Exact and Approximate Solutions of Nonlinear PDEs: The Role of the Simplest Equation. \emph{Commun. Nonlinear Sci. Numer. Simulat.} \textbf{16} (2011), 4215 -- 4231.
\bibitem{v92a}
	MARTINOV N., N. VITANOV. On Some Solutions of Two-Dimensional
	 Sine-Gordon 		Equation. \emph{ Journal of Physics A: Mathematical and General} \textbf{25} (1992), L419 -- L426.
\bibitem{v92b}
	MARTINOV N, N. VITANOV. Running Wave Solutions of the Two-Dimensional Sine-Gordon Equation. \emph{Journal of Physics A: Mathematical and General} \textbf{25} (1992), 3609 -- 3613.
\bibitem{v94}
	MARTINOV N. K., N. K. VITANOV. New Class of Running-Wave Solutions of the (2+1)-Dimensional Sine-Gordon Equation. \emph{Journal of Physics A: 
	Mathematical and General} \textbf{27} (1994), 4611 -- 4618.
\bibitem{vit96b}
	VITANOV N. K., N. K. MARTINOV. On the Solitary Waves in the Sine-Gordon Model of the Two-Dimensional Josephson Junction. \emph{Z. Phys. B} \textbf{100} (1996), 129 --135. 
\bibitem{b1}
	BISWAS A., S. KONAR. Introduction to non-Kerr Law Optical Solitons. 
	Chapman and Hall/CRC, London, 2006.
\bibitem{b2}
	BISWAS A., D. MILOVIC, M. J. EDWARDS. Mathematical Theory of Dispersion-
	Managed Optical Solitons. Springer, Berlin, 2010.
\bibitem{p3}
	PEREGRINE D. H. Water Waves, Nonlinear Schr{\"o}dinger Equations and 
	Their Solutions. \emph{J. Austral. Math. Soc. B}, \textbf{25} (1983), 
\bibitem{p7}
	ZHOU Q., D. YAO, S. DING, Y. ZHANG, F. CHEN, F. CHEN, X. LIU.
	Spatial Optical Solitons in Fifth Order and Seventh Order Weakly
	Nonlocal Nonlinear Media. \emph{Optik}, \textbf{124},  5683 -- 5686.
\end{thebibliography}
\end{document}